\documentclass{emulateapj} 
\usepackage{apjfonts}

\newcommand{\nc}{\newcommand}
\nc{\Porb}{$P_{\rm orb}$\,}
\nc{\Teff}{$T_{\rm eff}$\,}
\nc{\logg}{log\,$g$\,}
\nc{\kms}{\,${\rm km\,s}^{-1}$\,}
\nc{\Msun}{$M_{\odot}\ $}
\nc{\Mcz}{$M_{CZ}\ $}
\nc{\vsini}{$v \sin i$\,} 
\nc{\vmic}{$v_{\rm mic}$}
\nc{\vrad}{$v_{\rm rad}$}
\nc{\ALi}{A$_{\rm Li}$\,}
\nc{\Ali}{A$_{\rm Li}$\,}
\nc{\LeeZinn}{\mathcal{L}}

\slugcomment{Printed 2009 September 24 in ApJ}

\shorttitle{Rotation in Metal-Poor Stars}
\shortauthors{Cort\'es et al. 2009}

\begin{document}


\title{An Overview of the Rotational Behavior of Metal--Poor Stars}


\author{C. Cort\'es\altaffilmark{1}, J. R. P. Silva\altaffilmark{2}, A. Recio-Blanco\altaffilmark{3}, M. Catelan\altaffilmark{4,5,6}, J. D. Do Nascimento Jr.\altaffilmark{1}, J. R. De Medeiros\altaffilmark{1}}
\affil{}


\altaffiltext{1}{Departamento de F\'{i}sica, Universidade Federal do Rio 
            Grande do Norte, 59072-970 Natal, RN, Brazil.}
\altaffiltext{2}{Departamento de F\'{i}sica, Universidade do Estado do Rio Grande do Norte,
                 Mossor\'o, RN., Brazil.}
\altaffiltext{3}{Observatoire de la C\^ote d'Azur, Nice, France.}
\altaffiltext{4}{Departamento de Astronom\'{i}a y Astrof\'{i}sica, Pontificia Universidad
               Cat\'olica de Chile, Santiago, Chile.}
\altaffiltext{5}{John Simon Guggenheim Memorial Foundation Fellow}
\altaffiltext{6}{On sabbatical leave at Departamento de F\'{i}sica, Universidade Federal do Rio 
            Grande do Norte, 59072-970 Natal, RN, Brazil}


\begin{abstract}
The present paper describes the behavior of the rotational velocity in 
metal--poor stars ($\rm{[Fe/H]} \leq -0.5$~dex)
in different evolutionary stages, based on \vsini values 
from the literature. Our sample is comprised of 
stars in the field and some Galactic globular clusters, including stars on 
the main sequence, the red giant branch (RGB), and the horizontal branch (HB). 
The metal--poor stars are, mainly, slow rotators, and their \vsini distribution along 
the HR diagram is quite homogeneous. Nevertheless, a few moderate to high values of 
\vsini are  found in stars located on the main sequence and on the HB. We show 
that the overall distribution of \vsini values is basically independent of 
metallicity for the stars in our sample. 
In particular, the fast-rotating main 
sequence stars in our sample present similar rotation rates as their metal-rich 
counterparts, suggesting that some of them may actually be fairly young, in spite
of their low metallicity, or else that at least some of them would be better 
classified as blue
straggler stars. We do not find significant evidence of evolution in 
\vsini values as a function of position on the RGB; in particular, we do not 
confirm previous suggestions that stars close to the RGB tip rotate faster than 
their less evolved counterparts. While the presence of fast rotators among moderately 
cool blue HB stars has been suggested to be due to angular momentum transport from a 
stellar core that has retained significant angular momentum during its prior evolution, 
we find that any such transport mechanisms must likely operate very fast as the star 
arrives on the zero-age HB (ZAHB), since we do not find a link between evolution off 
the ZAHB and \vsini values. 

We present an extensive tabulation of all quantities discussed in this paper, 
including rotation velocities, temperatures, gravities, and metallicities [Fe/H], 
as well as broadband magnitudes and colors.
\end{abstract}




\section{Introduction}\label{sec:INTRO}

Rotation has long been an important factor affecting stellar evolution that 
has been largely ignored, especially when dealing with unresolved stellar 
populations, since the interplay between rotation and evolution is very difficult 
to accurately establish. Still, knowledge of stellar angular momentum evolution 
and its influence on a star's evolutionary history is clearly crucial to properly 
understand the evolution of stars. In spite of its importance 
in stellar astrophysics, the influence of rotation upon stellar evolution has  
not been properly established yet, and there have been relatively few studies dedicated 
to this subject. Recently, the new space telescopes dedicated to asteroseismology 
and the search for extra-solar planets, such as CoRoT and Kepler, have opened the 
possibility to determine rotation periods for large samples of stars in the solar 
neighborhood covering all main evolutionary stages. These observations, and the 
information that we can derive therefrom, can give us keys for the study of the 
angular momentum evolution and their effects on the stellar life~-- and, as a main 
result, the Sun's evolution. 

On the other hand, during the last two decades several studies have been aimed at 
describing rotation in metal-poor stars (e.g., Peterson 1983; Peterson 1985a,b; 
Peterson et al. 1995; Cohen \& McCarthy 1997; Behr et al. 2000a,b; Kinman et al. 
2000; Behr 2003a,b; Carney et al. 2003, Recio-Blanco et al. 2002, 2004; De Medeiros 
et al. 2006; Carney et al. 2008). While these studies have led to a large list of 
high-precision \vsini measurements for metal-poor stars, a comprehensive study of the 
rotational behavior of metal-poor stars based on these data has not been performed yet. 
Since metal-poor stars are mainly members of the oldest stellar populations, 
such a dataset may help shed light on the evolution of the angular momentum over a 
large range of evolutionary stages for low-mass stars. 

The distributions of \vsini in different evolutionary stages, from the main sequence (MS) 
to the red giant branch (RGB), as derived in these previous studies, show that metal-poor stars 
present, essentially, low \vsini values. Nevertheless, the stars in the horizontal branch 
(HB) do not show the same behavior and an enhanced rotation in these core helium-burning stars 
has been reported 
in several Galactic globular clusters (GCs) (e.g., Peterson et al. 1995; Cohen \& McCarthy 1997; 
Behr et al. 2000a,b; Recio-Blanco et al. 2002, 2004). The stars in these dense environments 
present a broad range of \vsini values, from several \kms to tens of \kms. This particular 
distribution is not explained as a natural evolution of the stellar angular momentum, since 
their ancestors, the RGB stars, present lower \vsini values. Field HB stars  
seem to have a similar distribution of \vsini values (Kinman et al. 2000; Behr 2003b; Carney 
et al. 2003), suggesting that the environment does not play a strong role in defining the 
observed spread in \vsini. Peterson et al. (1996) reported that the RR Lyrae stars, the variable stars 
located on the HB, present an upper limit of \vsini $\sim10$ \kms , indicating that these stars 
are also slow rotators and do not show the spread in \vsini found among other HB 
stars.

Soker (1998) and  Soker \& Harpaz (2000) have suggested 
that the \vsini distribution along the HB may bear the direct imprint of angular momentum 
transfer (from small-mass companions) during their previous RGB evolution. 
In particular, they have suggested that 
the high \vsini values found in several HB stars may be due to angular momentum transfer 
from a stellar or planetary companion, whose engulfment may have led to a spin-up of the 
primary star when the latter's external layers were expanded during the RGB phase. 
The added centrifugal acceleration to the RGB star's outermost layers could accordingly 
lead to an increase in the amount of mass lost by the RGB stars, thus leading to the 
formation of HB stars of lower mass, i.e., blue HB stars. Recently, Silvotti et al. (2007) found a giant 
planet orbiting V391 Pegasi, an extreme (blue) HB star, which lends some support to the  
planet engulfment scenario. However, no planets have hitherto been found in GCs, despite 
the fact that several surveys have been dedicated to identify planetary systems in GCs
(Gilliland et al. 2000; Weldrake et al. 2005). Otherwise, the distributions of metallicity for stars with planets 
show a bias for metal-poor stars, due to the fact that the majority of detected 
planets have as  host metal-rich stars (Santos et al. 2003; Santos et al. 2004; Fischer \& Valenti 2005). 
Other studies have argued that 
the \vsini values found in HB stars may be a consequence of complex angular momentum transfer 
mechanisms operating between the external layers and the degenerate cores of the RGB stars 
(e.g., Pinsonneault et al. 1991; Sills \& Pinsonneault 2000).

	Mengel \& Gross (1976) showed that a rapidly spinning RGB core may lead to a delay in 
	the onset of the He flash, and thus to an extension of the RGB phase towards higher luminosities 
	and lower temperatures~-- which could thus also lead to an increased mass loss and hence 
	to bluer HB stars (see also Sweigart 1997a,b).  
    Stellar rotation has thus been pointed out as a possible contributor to the 
	so-called ``second parameter'' phenomenon~-- namely, the presence of 
	GCs with similar metallicities but different color distributions along the HB
	(see Catelan 2009, for a recent review). 
	
The ``Grudahl jump" discontinuity (Grundahl et al. 1999), characterized by overluminous 
stars in blue passbands, most notably in Str\"omgren \textit{u} and Johnson $U$, is also 
reflected upon the values of \vsini found in blue HB stars  (Recio-Blanco 
et al. 2002). 
The overluminosity in $u$ 
was shown by Grundahl et al. to be due to the fact that HB stars with \Teff $> 11, 500$~K 
are strongly affected by radiative levitation and gravitational settling. The ensuing 
stellar winds (Vink \& Cassisi 2002) and strong chemical gradients (Sills \& Pinsonneault 
2000) could help spin down these stars. Interestingly, Recio-Blanco 
et al. (2002) suggested that there is no evidence for a link between evolution away 
from the zero age HB (ZAHB) and the \vsini values in HB stars.
	
The main aim of the present study is to carry out an analysis of the evolutionary behavior 
of rotation velocity in metal-poor stars, in the field and in Galactic GCs alike, 
based on a thorough compilation of \vsini measurements from the literature. 
The stars in our sample are widely distributed across the HR diagram, from  
the main sequence to later evolutionary stages, such as the RGB and the HB. The paper is 
structured as follows:  in \S\ref{sec:Sample} the main characteristics of the working sample 
are described. The main results are  presented in \S\ref{sec:Results}. Our conclusions 
are drawn in \S\ref{sec:Conclusions}. All data used in this work are presented 
in Tables~\ref{tab:FieldStars} and \ref{tab:GCStars}.

\section{The working sample}\label{sec:Sample}
For the purpose of this work we have compiled the \vsini values available in 
the literature for stars with $\rm{[Fe/H]} \leq -0.5$~dex. 
The stars are located in the field and in some Galactic GCs.
The sample of field stars was obtained from Kinman et 
al. (2000), Behr (2003b), Carney et al. (2003, 2008), and De Medeiros et al. (2006). These
stars are located in the MS, RGB, and HB phases.
On the other hand, the sample of stars in GCs is restricted to HB stars.
The data were obtained from Peterson (1985), Peterson et al. 
(1995), Cohen \& McCarthy (1997), Behr (2003a), and Recio-Blanco et al (2004). Several 
blue HB stars in metal-poor GCs present $\rm{[Fe/H]} > -0.5$ as a consequence of radiative 
levitation; we will discuss this further in \S\ref{subsec:HB}. 
Some relevant properties of the selected stellar sample are indicated 
in Table~\ref{tab:Authors}. 


\begin{deluxetable*}{lrcrr}
\tabletypesize{\scriptsize}
\tablecaption{Data Characteristics and References\label{tab:Authors}}
\tablewidth{0pt}
\tablehead{
\colhead{Author} & \colhead{\# stars} & \colhead{Method used} & 
\colhead{Resolution} &\colhead{$\langle{\rm Error}\rangle_{v \sin i}$}
}
\startdata
Perterson (1985)& 9 & Profile fitting & $23,\!000$ & 4.0 \kms \\
Peterson et al. (1995)& 63 & O~I triplet & $20,\!000$ & 3.2-6.0 \kms\\
Cohen \& McCarthy (1997) & 5 & Profile fitting & $36,\!000$ & 3.8 \kms  \\
Kinman et al. (2000)  & 28 & Gaussian fitting & $15,\!000$ & \nodata  \\ 
                    & & & $30,\!000$ & \nodata \\
                    & & & $40,\!000$ & \nodata \\
Recio-Blanco et al. (2002)& 63 & Cross-correlation & $40,\!000$ & 5.0 \kms  \\
Behr (2003a)& 74 & Profile fitting & $36,\!000$ & 3.0 \kms  \\
 & & & $45,\!000$ & \\
Behr (2003b) & 90 & Profile fitting & $45,\!000$ & 3.0 \kms   \\
 & & & $60,\!000$ & \\
Carney et al. (2003) & 80 & Profile fitting & $35,\!000$ & 0.5-2.0 \kms\\
De Medeiros (2006) & 99 &  Profile fitting & $48,\!000$  & 2.0 \kms\\
 & & & $50,\!000$ & \\
Carney et al. (2008)& 19 & Fourier decomposition& $120,\!000$ & 1.0 \kms  \\
\enddata
\end{deluxetable*}

Nine HB stars considered in Peterson (1985) belong to the GC M~4 (NGC~6121), which
was observed using the echelle spectrograph with a spectral 
resolution $R\sim23,\!000$ at the Multiple Mirror Telescope Observatory. The \vsini 
measurements were carried out using some Mg lines and a Fe~II line, with typical errors 
in the \vsini values of 4~\kms. 

The \vsini values listed by Peterson et al. (1995) 
were derived using the O~I line triplet at 7771-7775~\AA. Their stellar sample 
contains HB stars in three GCs: M~3 (NGC~5272), M~13 (NGC~6205), and NGC~288. The
spectra were obtained at medium resolution  ($R\sim20,\!000$), and were collected with the fiber-fed system 
``Nessie" with an echelle spectrograph at the 4~m telescope at Kitt Peak Observatory/NOAO. The typical 
errors in the \vsini measurements are 5.0, 3.2, and 6.0~\kms for the stars in M~3, M~13, and NGC~288, 
respectively. 

Five blue HB stars in M~92 (NGC~6341) were analyzed by Cohen \& McCarthy (1997). They used the HIRES 
spectrograph on the Keck I telescope atop Mauna Kea to obtain high resolution spectra 
for these stars ($R\sim36,\!000$). Two methods were used to derive \vsini. 
The first was a comparison of Gaussian line fits for isolated and unblended stars 
with similar profile measurements for the arc emission lines, varying the strength of 
the rotational broadening. The second one was a comparison between the profiles of 
strong absorption lines of Fe and Ti with the profile of a single Th-Ar arc emission 
line. We decided to use only the data based on the latter method, since it appears to  
be more reliable for fast rotating stars. The average error in the \vsini measurements 
is 3.8~\kms. 
	 
The \vsini values of blue HB stars measured by Kinman et al. (2000) were obtained from  medium- 
($R\sim15,\!000$) and high-resolution spectra ($R\sim30,\!000$ and $R\sim40,\!000$), collected 
at the Kitt Peak 0.9~m coud\'e feed spectrograph and the CAT+CES (1.4~m Coud\'e Auxiliary Telescope + Coud\'e Echelle Spectograph) combination at La Silla, Chile.  
In this work the rotational velocities were obtained using the FWHM of the Mg~II line 
($\lambda4481$) and  the fit of the profile Mg~II lines of the observed spectra with 
the synthetic one. When both mesuarements  are available for a stars, we obtained the \vsini from the
average between  both values. These data may have important uncertainties, but unfortunately the associated 
measurement errors are not available.

The Recio-Blanco et al. (2002, 2004) data refer to four metal-poor Galactic GCs, namely NGC~2808, 
M~15 (NGC~7078), M~79 (NGC~1904), and M~80 (NGC~6093). The stellar spectra were obtained with high 
resolution ($R\sim40,\!000$) using UVES at the Kueyen-VLT. The values of \vsini for each star were 
obtained using the procedure described in Tonry \& Davis (1979) and Melo, Pasquini, \& De Medeiros 
(2001). The typical error in the computed \vsini is around 5~\kms. Most stars 
contained in this work do not have associated [Fe/H] measurements, but we were able to 
incorporate [Fe/H] values for some of the stars in M~79 based on the Fabian et al. (2005) study.

The stars in Behr (2003a) were observed with the HIRES-Keck spectrograph using a spectral resolution 
$R\sim45,\!000$ and $R\sim36,\!000$. To compute \vsini he used the minimum value of $\chi^{2}$ 
for the fit between the metal absorption lines and Kurucz synthetic spectra. The stars belong 
to six Galactic GCs: NGC~288, M~3, M~13, M~15, M~68 (NGC~4590), and M~92.  
The typical error in \vsini is 3.0~\kms. Behr's (2003b) sample, in turn, contains 
field  stars only. The stars were observed at high resolution ($R\sim60,\!000$ 
and $R\sim45,\!000$) using the Cassegrain echelle spectrograph on the McDonald Observatory 2.1 m 
Otto Struve Telescope and the HIRES-Keck spectrograph. The \vsini values were calculated using the 
same method as in Behr (2003a). The maximum error in the determination of \vsini is 3.0~\kms. 

Carney et al. (2003) used field RGB and red HB stars, which were observed with the 1.5~m Wyeth 
reflector at the Oak Ridge Observatory in Harvard, Massachusetts. They also used the 1.5~m Tillinghats 
reflector and the Multiple Mirror Telescope atop Mount Hopkins, in Arizona. The \vsini values were derived 
from the comparison of observed rotational broading with synthetic spectra, and the typical error 
in \vsini measurements ranges from 0.5 to 2.0~\kms. 
	
	The working sample in De Medeiros et al. (2006) is comprised of  metal-poor field stars in three 
	 evolutionary stages, from the MS to the HB. The 	spectra were observed with high resolution 
	 ($R\sim48,\!000$ and $R\sim50,\!000$) at the FEROS spectrometer 
	mounted on the ESO 1.5~m telescope, together with the CORALIE spectrometer mounted on the Euler 
	Swiss 1.2~m telescope, both at La Silla, Chile. The \vsini values of the stellar sample were 
	calculated by fitting the observed spectra with a synthetic one. They used metallic lines to 
	compute \vsini. Typical error on the \vsini measurements for their working sample is around 2~\kms. 
	
	The 
	data compiled in Carney et al. (2008) corresponds to 12 metal-poor field RGB stars and 7 metal-poor 
	field red HB stars, which were observed using the Gecko spectrograph at the Canada-France-Hawaii 
	Telescope (CFHT) with a typical $R\sim120,\!000$. In this work the authors identified the \vsini 
	and the macroturbulence component in the Doppler broadening in the line profiles using a Fourier 
	analysis. The typical error in the \vsini measurements is $\sim1.0$~\kms.
	
Some HB stars in GCs do not have measurements of their surface gravities and iron abundances. 
For some stars in the GCs M~3, M~15 and NGC~288 we compiled the surface gravities from Crocker et al. (1988) 
and Moehler et al. (1995). In order to have an estimation of the surface gravities for  stars without
any measurements in M~3, M~13, M~15, M~79, M~80, M~92, NGC~288,
 and NGC~2808, we used empirical (though approximate) 
 relations between \Teff and $\log(g)$. More specifically, for the GCs M~3, M~13, 
 M~79, and M~80, we used eq.~(2) in Fabian et al. (2005),  whereas for the GCs NGC~288 and NGC~2808 we used 
 eq.~(1) in Pace et al. (2006). For M~15 we used the values of $\log(g)$ found in Behr (2003a) to interpolate (or extrapolate) 
 the surface gravities for stars in M~15. Otherwise, for those stars without  $\rm{[Fe/H]}$ measurements, 
 we assumed that the amount of heavy elements is equal to the characteristic abundances in the GC, which are
 presented in Table~\ref{tab:GCparam} (Harris 1996).  Obviously we can only assume this for stars with \Teff$\leq 11,500$~K.

\begin{deluxetable*}{lcccc}
\tabletypesize{\scriptsize}
\tablecaption{Chemical and Photometrical Parameters for  Galactic GCs\label{tab:GCparam}}
\tablewidth{0pt}
\tablehead{
\colhead{Cluster} & \colhead{$\rm{[Fe/H]}$} & \colhead{$M_{V}$} & 
\colhead{$(m-M)_{V}$} & \colhead{$E(\bv)$}
}
\startdata
 NGC~288         & -1.24 & -6.74 & 14.83 & 0.03   \\
 NGC~2808        & -1.15 & -9.39 & 15.59 & 0.22   \\  
 M~3 (NGC~5272)  & -1.57 & -8.93 & 15.12 & 0.01   \\
 M~4 (NGC~6121)  & -1.20 & -7.20 & 12.83 & 0.36   \\
 M~13 (NGC~6205) & -1.54 & -8.70 & 14.48 & 0.02   \\
 M~15 (NGC~7078) & -2.26 & -9.17 & 15.37 & 0.10   \\              
 M~68 (NGC~4590) & -2.02 & -7.35 & 15.19 & 0.05   \\
 M~79 (NGC~1904) & -1.57 & -7.86 & 15.59 & 0.01   \\
 M~80 (NGC~6093) & -1.75 & -8.23 & 15.56 & 0.18   \\ 
 M~92 (NGC~6341) & -2.28 & -8.20 & 14.64 & 0.02   \\
\enddata
\end{deluxetable*}

In Tables~\ref{tab:FieldStars} and \ref{tab:GCStars} we summarize the data for the field 
and GC stars analyzed in this paper, respectively. As we can see, several stars were
analyzed in more than one study. For these stars we averaged the values of their 
physical and chemical parameters, as well as their measured  rotational velocities, and we used 
these averaged values in our analysis.
In Table~\ref{tab:FieldStars} we  also present some photometrical quantities for our field stars.  
The $V$ magnitude and $(\bv)$ color were compiled from SIMBAD\footnote{http://simbad.u-strasbg.fr/simbad/}, 
and the $(V-I)$ color comes from the Hipparcos catalog (Perryman et al. 1997 and references therein). 
References to the \vsini measurement sources are also included in this table.  
In Table~\ref{tab:GCStars} we present the apparent magnitude $V$ and color index $(\bv)$.
For the stars in M~79 the $V$ values are based on measurements using the \textit{y} passband of 
the Str\"omgren system, since $V$ and $y$ magnitudes are well known to be basically identical 
(Clem et al. 2004).  We calculated the absolute magnitude $M_{V}$ and the unreddened color $(\bv)_0$ 
provided in Table~\ref{tab:GCStars}  using the  distance modulus and 
reddening values compiled in Harris (1996) for each GC (see Table~\ref{tab:GCparam}).

%
%
%
%
%
%

In order to improve the analysis, we  divided our stellar sample into different evolutionary stages, leading to 
a total of 51, 131, and 277 stars on the main sequence, the sub-giant branch (SGB) and RGB, and the HB, respectively. We note that 
in the RGB sample there are 14 confirmed binary systems (see table 5 in Carney et al. 2003 for details). The HB 
group was divided in two subgroups, the field HB group with 70 stars, and the HB in GCs group with 207 stars. 
Let us stress that there is an important difference between the resulting field and GC HB samples: while 
the field HB sample is comprised of red and blue HB stars (though with \Teff $< 11,500$~K), the GC HB sample 
is entirely comprised of blue HB stars. 

%

\begin{figure*}
\includegraphics*[width=6.5in]{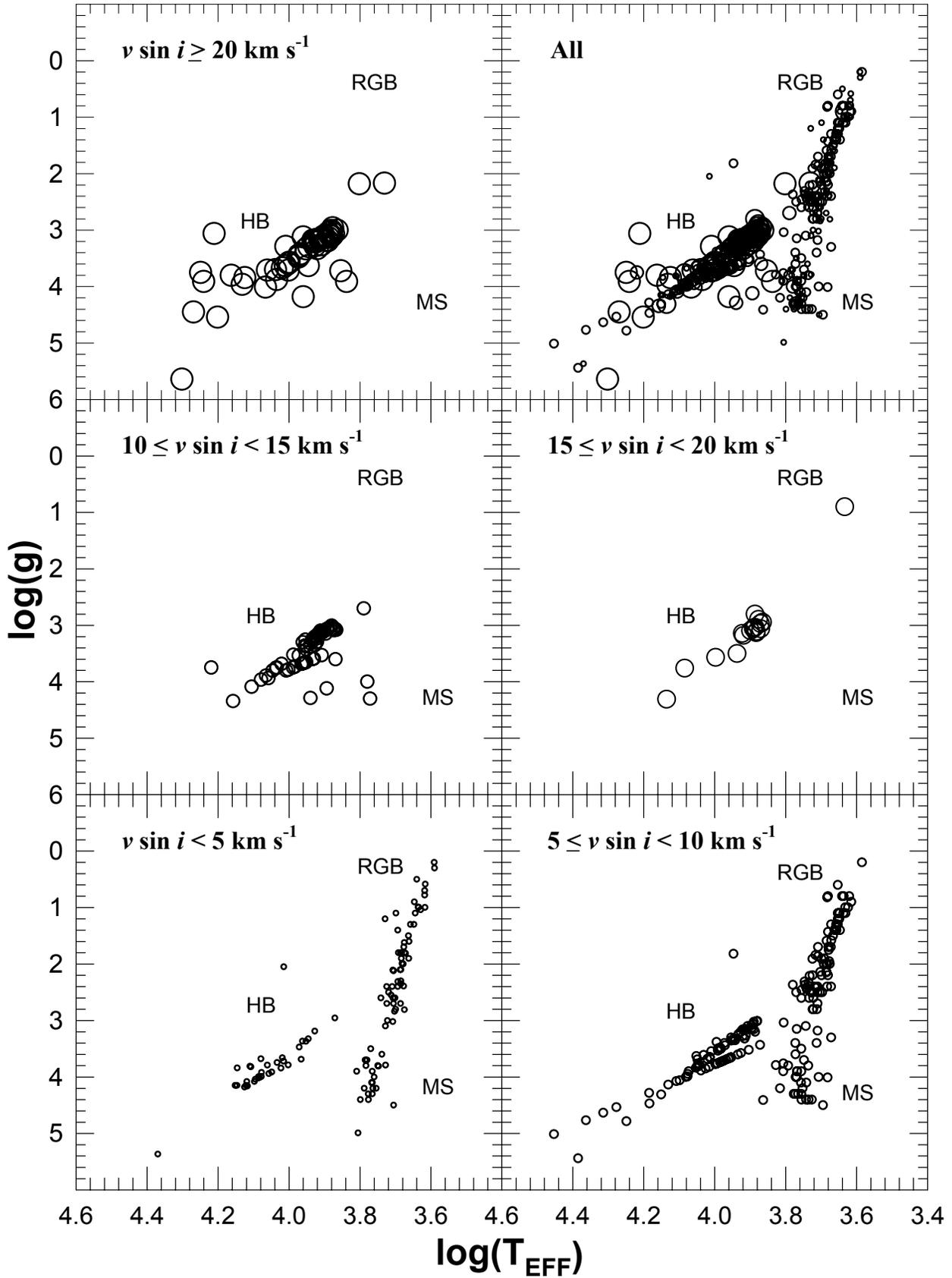}
\caption{The distribution of \vsini on the HR diagram for the stars in our sample. The panels 
            represent different \vsini intervals, as indicated. 
            The main evolutionary stages are also schematically indicated.
            Note that slow rotators can be found throughout the HR diagram, whereas the 
            fast rotators are found almost exclusively over a restricted range in HB
			temperatures. }
              \label{fig:Fig1}
\end{figure*}
%



\section{Results and Discussion}\label{sec:Results}

Figure~\ref{fig:Fig1} shows the distribution of \vsini along the HR diagram for metal-poor stars. 
This figure shows that most of these stars present low values of \vsini, and that 
these slow rotators are located along all evolutionary stages, from the MS to the HB.
There are several stars with high \vsini, and those are mainly located on the HB. However,
we can identify some fast rotators that are not on the HB, but which were rather  
catalogued as MS stars by Behr (2003b).   
We will discuss these stars in more detail  in \S\ref{subsec:MS}. Note
also that the RGB stars present almost exclusively low values of \vsini, with 
a single exception.


\begin{figure}

  \includegraphics*[width=3.2in]{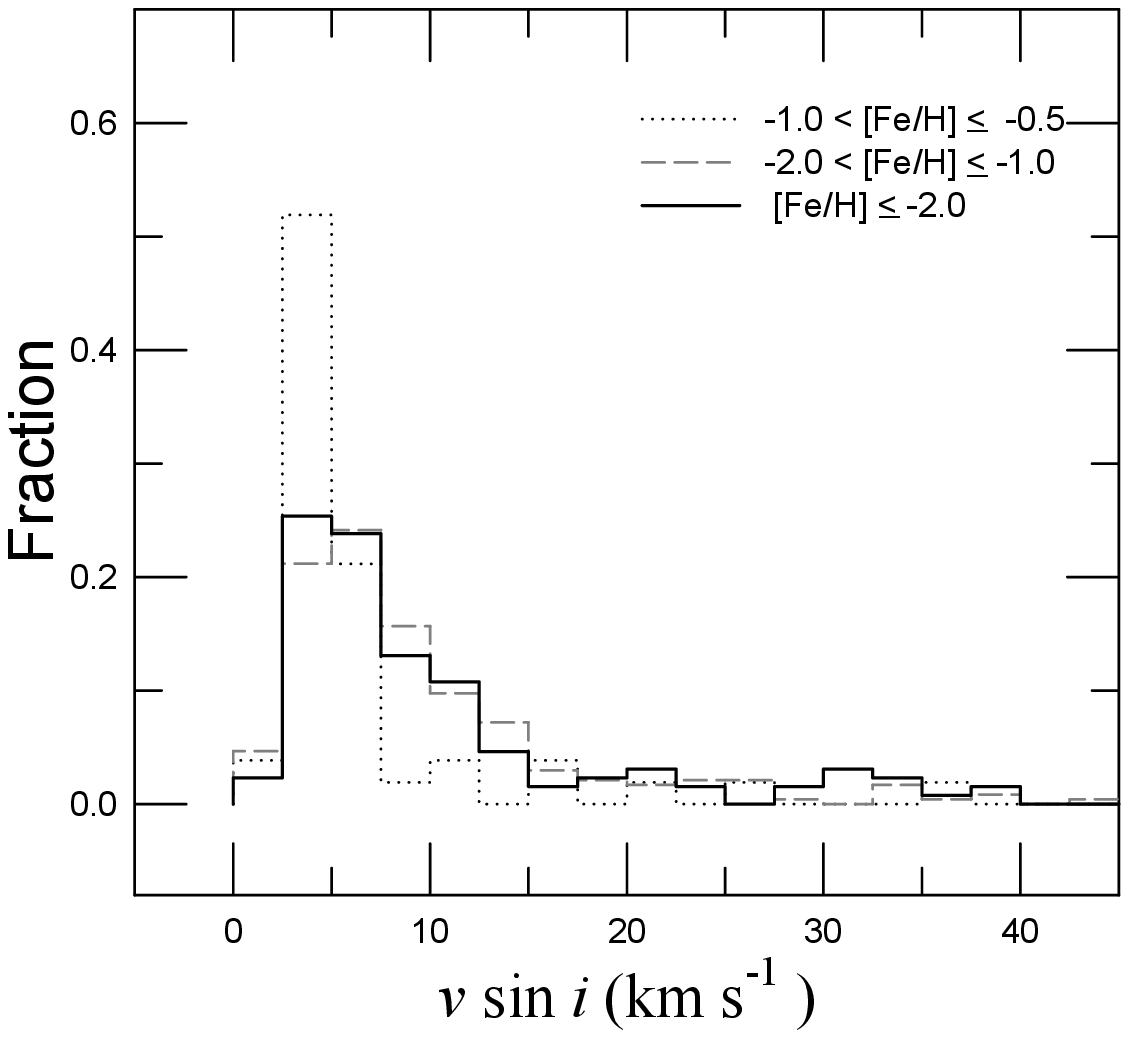}
    \caption{Histograms of \vsini as a function of  
            $\rm{[Fe/H]}$ for the stars in our stellar sample.  
            Groups FE1, FE2, and FE3 are shown (see text). The \vsini 
			distribution does not appear to present a significant
            dependence on metallicity.}
            \label{fig:Fig2}
    \end{figure}

   \begin{figure}
   \includegraphics*[width=3.2in]{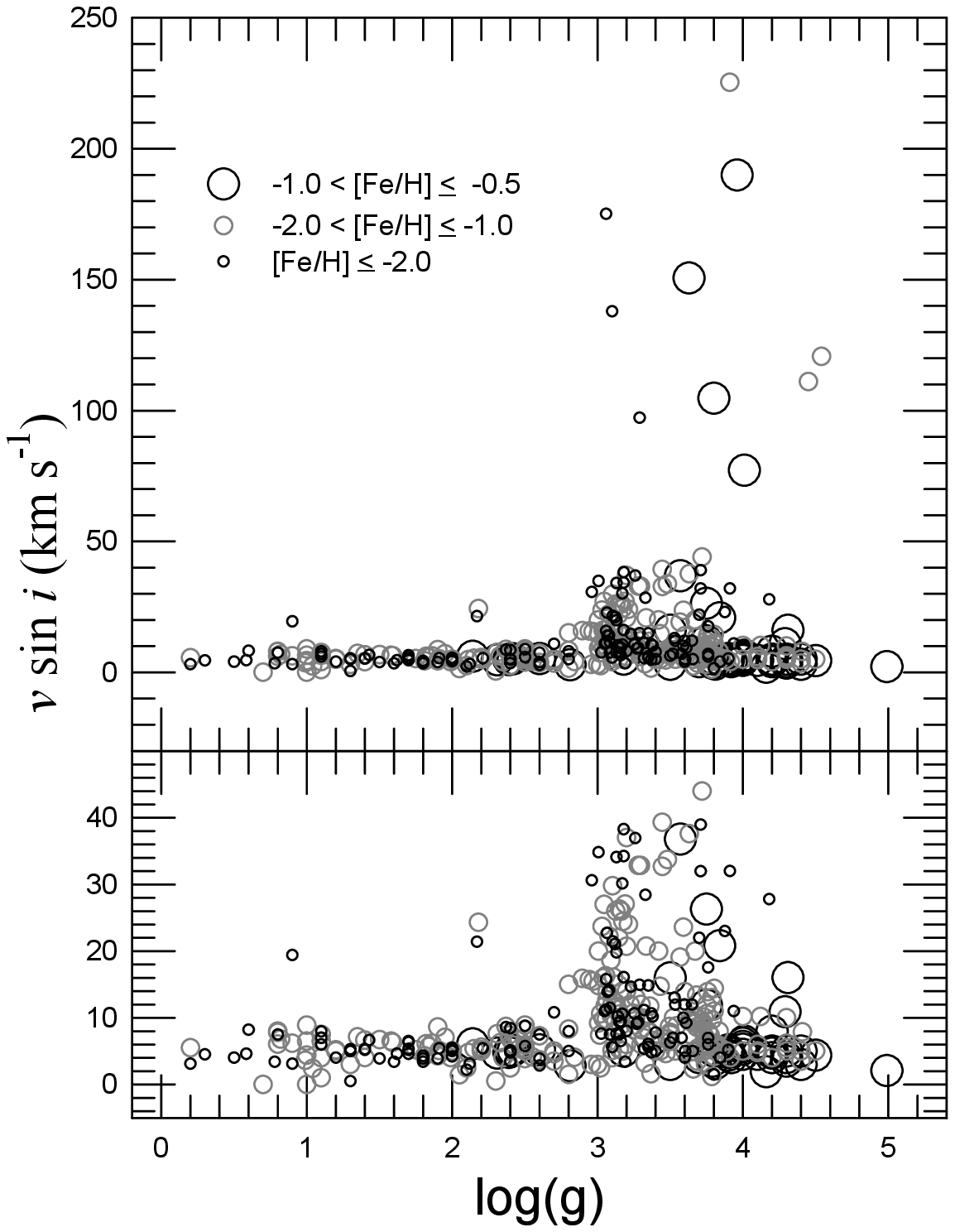}
   \caption{\vsini distribution as a function of surface gravity  $\log (g)$ for 
           different metallicity intervals. The lower panel represents a zoom 
		   around the low-\vsini region shown in the upper plot.
		   The \vsini distribution for a given $\log (g)$
           does not appear to show a significant dependence on metallicity.
           }\label{fig:Fig3}
    \end{figure}

We have separated our sample into three different groups, where the 
stars are organized by iron abundance. We made this separation in order 
to identify possible systematic differences in the behavior of \vsini for the different metallicities 
considered here. The first group, FE1, is formed by 52 stars with $-1.0 < \rm{[Fe/H]} \leq -0.5$; 
the second group, FE2, presents 236 stars with $-2.0 \leq \rm{[Fe/H]} < -1.0$; and the last group, 
FE3, presents 130 stars with $\rm{[Fe/H]}\leq-2.0$. There  are also 19 HB stars with unknown metallicities
and 22 metal-poor HB stars affected by radiative levitation ($-0.5<\rm{[Fe/H]} < +0.75$). 
We show the histograms of \vsini of groups  FE1, FE2, and FE3 in Figure~2, where we can see 
that the distributions are very similar. In particular, the range in \vsini values 
is closely the same for all groups. Also, the fraction of stars contained in the interval
\textbf{$0.0 \leq $\vsini$ < 15.0$} \kms is 0.83, 0.84, and 0.80 for the groups FE1, 
FE2, and FE3, respectively. 
We stress that groups FE2 and FE3 present particularly similar \vsini distributions, with the
histograms presenting closely the same shapes and peak locations; note that these two groups 
contain most of the stars in our sample. On the other hand, 
the distribution of \vsini for group FE1 presents a higher peak at $2.5\leq $\vsini$ < 5.0$ \kms
than do the other groups: indeed, the fraction of stars in this \vsini 
interval is 0.56, 0.26, and 0.28 for FE1, FE2, and FE3, respectively. 
Figure~3 shows the values 
of surface gravities $\log(g)$ and \vsini for the same groups. This figure reveals 
that the differences in the \vsini distributions are related to the number of stars in the different 
evolutionary stages, the stars in the FE2 and FE3 groups covering all values of $\log(g)$ and being located 
in all evolutionary stages, whereas most of the stars contained in the FE1 group present high values of 
$\log(g)$ and are mainly located on the MS. However, we can note that there are a few FE1 stars located 
on the RGB and the HB, and that these evolved stars present the same rotational behavior as do the stars 
in groups FE2 and FE3 at a similar evolutionary stage. This suggests that metallicity is not an 
important parameter defining stellar rotation~-- or, at least, we do not identify important 
differences in the rotational behavior between the metallicity groups considered here. Certainly, more 
observations are required to analyze the behavior of the rotation in RGB and HB stars, especially over 
the metallicity range $-1.0 < \rm{[Fe/H]} \leq-0.5$, in order to corroborate the suggested mild 
dependence of  \vsini with $\rm{[Fe/H]}$. 

Note, on the other hand, that there 
are some observational biases in the available field star samples. 
Mainly, the observed stars with measurements of \vsini tend to be bright in the sky, and their parallaxes 
show that they are indeed found mainly in the solar neighborhood~-- which can be understood because,  
in order to obtain reliable measurements of \vsini, high signal-to-noise ratio and medium-to-high spectral 
resolution are required. The low apparent magnitude is an observational constraint which can in principle 
also cause bias. In fact, the distribution of metallicities can be affected by choosing stars with low magnitudes; 
note, in particular, that the stars in our sample have mostly $V\leq12$, and  there are only two 
stars with $V\sim14$. Although several works have been developed to search for differences in the 
rotational behavior as a function of position in the Milky Way (e.g., De Medeiros et al. 
2000), there is no strong evidence showing important differences in the stellar rotation as a function 
of the Galactic latitude or the distance from the Galactic center. However, we stress the fact that 
the studies that have been carried out so far are focused on relatively bright stars. 
When future studies consider a representative sample both in magnitudes and Galactic positions, 
we should be in a position to derive stronger constraints on the behavior of \vsini as a function 
of position in our galaxy. We thus warn the reader that our study may suffer from such a source of bias, 
which however will only be possible to reliably quantify when significantly enlarged samples become available.

   \begin{figure*}
   \centering
   \includegraphics*[width=7.0in]{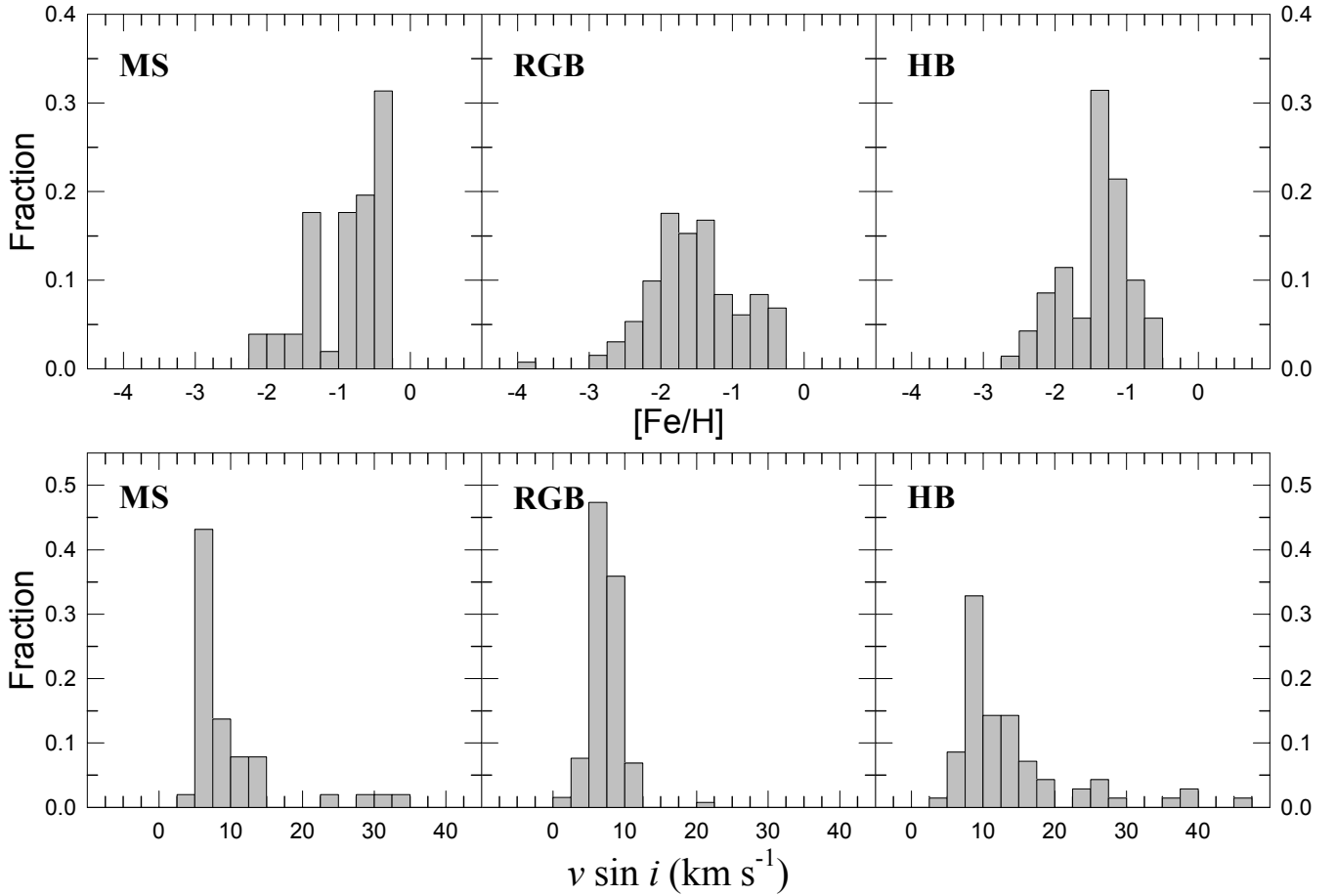}
   \caption{Histograms of $\rm{[Fe/H]}$ (\textit{upper panel}) and \vsini (\textit{lower panel}) for field stars
    on the MS, the RGB, and the HB (from \textit{left} to \textit{right}). 
	HB stars whose abundances are affected by radiative levitation are not included. 
	There are clearly strong differences between the \vsini distributions for RGB and HB stars.}
   \label{Histo}%
    \end{figure*}
%


\subsection{Main Sequence and Turn-Off Stars}\label{subsec:MS}
The distribution of $\rm{[Fe/H]}$ for MS stars in the present sample 
is shown in Figure~\ref{Histo} (\textit{upper left panel}). 
There are 51 MS and turn-off stars in our working sample, distributed across 
the three metallicity groups FE1, FE2, and FE3 descrived above according to 
the following percentages: 50.1$\%$ (26 stars), 41.1$\%$ (21 stars) 
and 7.8$\%$ (4 stars), respectively. 
Figure~\ref{fig:Fig3} shows that the MS stars in the different 
metallicity groups  present similar rotational behavior. This is 
especially clear for groups FE1 and FE2.

\begin{figure}
   \centering
    \includegraphics*[width=3.2in]{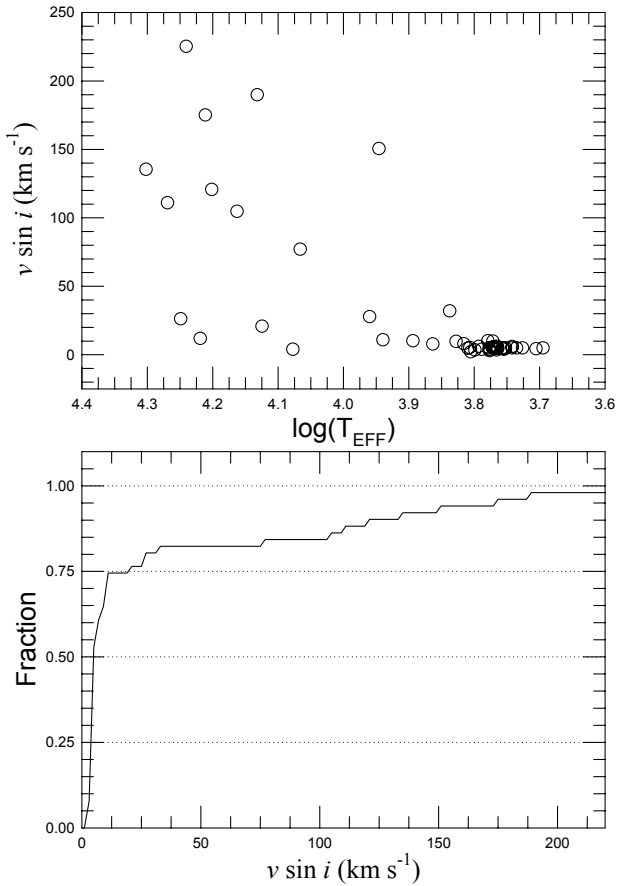}
   \caption{\textit{Upper panel:} \vsini values for the  field 
	main sequence stars as a function of  \Teff. \textit{Lower panel:} cumulative
	\vsini distribution for  MS stars.}
              \label{MS}%
    \end{figure}

The \vsini distribution  of our metal-poor MS stars 
shows that most of the hydrogen core burning stars are slow rotators 
(see Fig.~\ref{Histo}, \textit{left lower panel}). However, we also note
 that there are a few stars with high \vsini values. Figure~\ref{MS} shows the projected 
 rotational velocity  \vsini  as a function of $\log(T_{\rm eff})$ 
 (\textit{upper panel}) and  the corresponding cumulative \vsini distribution  
 (\textit{lower panel}). 
 The \textit{upper panel} in this figure shows that 
there is a clear  relation between \vsini and \Teff, which may also be partly understood 
as a relation between \vsini and stellar mass. We can see clearly that
 the hotter (or more massive)  MS stars present a spread in \vsini values, with values 
 reaching up to a few hundred \kms.
The cooler (or less massive) MS stars, on the other hand, present low \vsini values only.  
More specifically, 74.5$\%$ (38 stars) of them  present $0.0<$\vsini$\leq15.0$\kms, 7.8.$\%$ (4 stars)  
 present $15.0<$\vsini$\leq50.0$\kms, and 17.7$\%$ (9 stars) present \vsini$>50.0$\kms.

Many studies have been conducted over the years in order to understand the rotational behavior of stars 
in the MS phase (e.g., Wilson 1966; Kraft 1967; Smith 1979; Soderblom 1983; Stauffer et al. 1984; 
Stauffer et al. 1985; Melo et al. 2001). These works found that the metal-rich MS stars can 
present high values of \vsini, but these values are primarily linked with the stellar ages and  
masses.   Skumanich (1972) and  Pace \& Pasquini (2004) found some observational laws to describe 
the  evolution of the rotation velocity in low-mass stars, where we can see that the evolved 
metal-rich MS stars present low values of \vsini. However, these studies 
were focused on metal-rich stars in the field and several open clusters, and  no similar work 
has ever been conducted in the metal-poor regime. In this context,  it is very important to verify
whether the empirical laws discovered by the quoted authors remain valid at low metallicity.

Considering that the stars with high \vsini values
all have $\log(T_{\rm eff})\geq 3.825$ (and are thus relatively massive), 
and assuming that the quoted observational laws describing the evolution 
of rotation velocities can be extrapolated to the 
metal-poor domain, this should imply that the fast rotating MS stars in our 
sample are predominantly young. Still, the possibility that these stars present
older ages is also worth discussing. In fact, their position in the HR 
diagram may lead one to question the evolutionary stage assigned to these 
fast rotators. In this sense, if these stars belonged to the HB, we should 
definitely expect to find the chemical patterns characteristic of radiative
levitation and gravitational diffusion, including supersolar [Fe/H] values  
for stars with \Teff$>11,500$ [$\log(T_{\rm eff})\geq 4.06$]. 
However, these stars are all metal-poor, thus rendering this possibility 
unlikely.

 Another possibility is that some of these stars could be hypervelocity 
stars ejected from the Galactic center, since high rotation velocities have 
recently been measured for some such stars (L\'opez-Morales \& Bonanos 2008). However, none of the 
stars in our sample appears to be a hypervelocity star. Perhaps a more realistic explanation 
could be that at least some of these stars are actually field blue straggler (BS) 
stars, particularly in view of the fact that at least some such stars are also 
fast rotators (e.g.,Ryan et al.2002). Such a possibility cannot be discarded, 
because the BSs present similar physical parameters as the hot MS stars in our 
sample.

On the other hand, for those MS stars with $\log(T_{\rm eff})\leq 3.825$ 
(or less massive), the \vsini values are low. We would like to compare
the values of \vsini for these stars with the values found in GCs. 
The average \vsini for field stars is 
$\langle$\vsini$\rangle=5.18\pm1.66$~\kms. 
Lucatello \& Gratton (2003) found in the 
GCs NGC~104 (47~Tuc), NGC~6397 and NGC~6752 an upper limit of $3.5\pm0.2$~\kms for 
the \vsini of turn-off stars (specifically, they found $4.0\pm0.4$~\kms, 
$3.1\pm0.3$~\kms, and $3.6\pm0.3$~\kms for 47~Tuc, NGC~6397, and NGC~6752,
respectively). For our sample, the mean value of 
\vsini is higher than the upper limit  found in the GCs. We note that 
this is not a strong evidence of differences  or similarities in the  
rotational behavior of stars in field and clusters, because of the paucity
of both samples. Nevertheless, if the Skumanich (1972)  
law can describe the rotation of the metal--poor stars, we note that the difference 
between the mean value found here and the mean value given by Lucatello \& 
Gratton could result from a difference in ages for the stars in the
field and in the GCs, since stars in GCs are generally older than those in the field. 
 On the other hand,the positions of the stars of Lucatello \& Gratton 
in the color-magnitude diagram 
show that the stars in NGC~6397 and NGC~6752 are on the turn-off (Gratton et 
al. 2001), while the stars in 47~Tuc are just above the turn-off point 
(Carretta et al. 2004). Interestingly, Melo et al. (2001) found that the stars just 
above the turn-off point in M~67 (NGC~2682)
present a reduction in rotation velocities, with the stars below the 
turn-off presenting \vsini values almost 50$\%$ higher. The possibility cannot 
be excluded that age effects may also be present, since it has been suggested 
that 47~Tuc may be slightly younger than NGC~6397 or NGC~6752 (e.g., Gratton et 
al. 2003).

\subsection{Sub-Giant and Red Giant Branch Stars}\label{subsec:RGB}

The \textit{central upper panel} in Figure~\ref{Histo} shows that 
the $\rm{[Fe/H]}$ distribution of the SGB and RGB stars considered in 
our stellar sample closely resembles a normal distribution. There are 131 stars 
in this stage, distributed into groups FE1, FE2, and FE3 in the following proportions, 
respectively: 15.3$\%$ (20 stars), 46.6$\%$ (61 stars), and
38.2$\%$ (50 stars). 

\begin{figure}
   \centering
    \includegraphics*[width=3.2in]{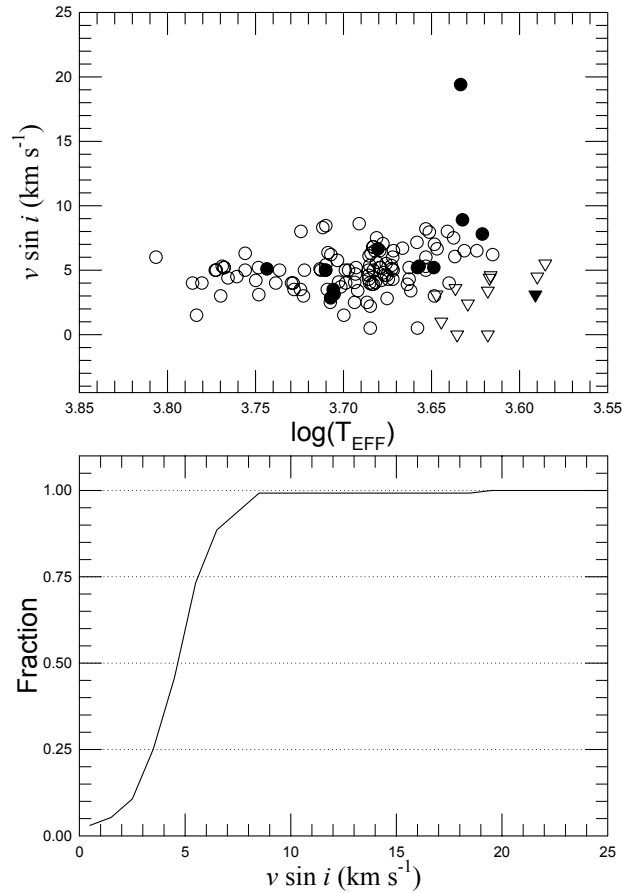}
  \caption{\textit{Upper panel:}  \vsini values for the  field 
	RGB stars as a function of  \Teff. 
	The open circles represent  single RGB stars, 
	whereas the filled black circles represent binary systems. 
	Binary systems present a similar behavior as  single 
	RGB stars. The stars of Carney et al. (2008) are also presented,
	with the open triangles representing  single RGB stars in their sample, 
	and the filled black triangles representing binary systems studied by these authors. 
	\textit{Lower panel:} cumulative \vsini distribution for the RGB stars.} 
              \label{rgb}%
    \end{figure}

The histogram of \vsini for SGB and RGB stars (see Fig.~\ref{Histo}, \textit{central 
lower panel}) shows that most of these stars with expanding photospheres are slow rotators.   
In  Figure~\ref{rgb} we show  \vsini as a function of effective 
temperature for these stars (\textit{upper panel}), where we have 
included both single stars (open circles) and stars in confirmed binary 
systems  (filled circles). Figure~\ref{rgb} also includes the recent 
measurements of \vsini in RGB stars by Carney et al. (2008), where we have included the 
single RGB stars (open inverted triangles) and the confirmed binary systems 
(filled inverted triangles). The cumulative \vsini distribution for all SGB and RGB stars 
is also presented in Figure~\ref{rgb} (\textit{lower panel}). Most SGB and RGB
stars present low \vsini values (\vsini$\leq15.0$\kms). The RGB stars in binary 
systems present similar rotational behavior, for a given \Teff, to the single 
RGB stars. There is a single binary system presenting high rotation (19.4 \kms), 
namely star $\rm{CD-37^{\mathrm{o}} 14010}$, which has an orbital eccentricity $e=0.058$ 
and an orbital period $T=62.55\,d$ (Carney et al. 2003). These orbital 
parameters suggest that this star is synchronized, thus explaining its enhanced
rotation.


Another important result is that the stars pass along the RGB with
\vsini values that seem to remain practically constant (Fig.~\ref{rgb}).   
This feature was also found in  stars in 
the field and open clusters (Pasquini et al. 2000; Melo et al. 2001). 
Naturally, if these stars conserve angular momentum,  the expansion in their 
radius should produce a reduction of the \vsini value 
at the surface. In addition, the rotation should slow down  due to the angular 
momentum that is lost as the star loses mass during its approach of the RGB tip. 
However, the RGB stars in our sample do not show this feature, i.e., 
they do not slow down as they approach the RGB tip~-- as also reported by Carney 
et al. (2008). As a matter of fact, if one considers solely the Carney et al. 
(2008) sample, one is led to conclude that the stars arriving at the tip of the 
RGB may even present an {\em enhanced} rotation (see inverted triangles in Fig.~\ref{rgb}). 
However, this enhancement does not become apparent when data from other studies 
are also incorporated, as can also be seen from Figure~\ref{rgb}. More data for 
stars close to the RGB tip are needed to conclusively settle this issue. 
 
The mean values  of \vsini are also calculated for the single stars and the binary 
systems, and we found that these values are $<$\vsini$>=4.79\pm1.75$ \kms and 
$<$\vsini$>=6.24\pm4.35$ \kms, respectively. 
When we do not take into account the $\rm{CD-37^{\mathrm{o}} 14010}$
star, the mean value of the binary system is $<$\vsini$>=5.14\pm1.89$ \kms. 
 If  the stars of Carney et al. (2008) are not considered, we obtain $<$\vsini$>=4.98\pm1.63$ \kms 
 and $<$\vsini$>=5.33\pm1.87$ \kms  
for the single stars and the binary systems (without the  star $\rm{CD-37^{\mathrm{o}} 14010}$), 
 respectively.

\subsection{Horizontal Branch Stars}\label{subsec:HB}

The \textit{right upper panel} in Figure~\ref{Histo} shows the distribution of  
$\rm{[Fe/H]}$ for the field HB stars. There are 70 stars in this stage in 
our sample. 5.7$\%$ (4 stars), 68.6$\%$ (48 stars), and 25.7$\%$ (18 stars) are the 
percentages of HB stars in groups FE1, FE2, and FE3, respectively.

The histogram of \vsini for field HB stars (see Fig.~\ref{Histo}, 
\textit{right lower panel}) shows that the helium core burning stars cannot 
be generally considered slow rotators, because they present a spread of \vsini values. 
In spite of the fact that HB stars present a narrow range in masses 
(from about 0.5 to 1.0 $M_{\odot}$), they show \vsini values up to several tens of \kms. 
This  characteristic cannot be straightforwardly explained in terms of canonical stellar 
evolution models, because their ancestors (RGB stars and low-mass MS stars), as we noted in 
previous sections, are mainly slow rotators. 

In order to observe the differences or similarities in the behavior of the rotation 
between HB stars in the field and GCs, we have compared the values for field stars 
with those for stars in the GCs NGC~288, NGC~2808, M~3, M~4,
M~13, M~15, M~68, M~79, M~80, and M~92. For those stars without individual
iron abundances and \Teff $< 11,500$ K, we assume the typical GC values  (see 
Table~\ref{tab:GCparam}). The HB stars in M~4 do not present \Teff mesuarements, but
their (\bv) colors point to \Teff $< 11,500$~K. 1.0$\%$ (2 stars), 51.2$\%$ (106 stars), 
and 28.0$\%$ (58 stars) are the percentanges  of stars in groups FE1, FE2 and FE3, 
respectively. 10.6$\%$ (22 stars) present enhanced metallicity ($\rm{[Fe/H]}>$-0.5), due
to radiative levitation, and 9.2$\%$ (19 stars) do not have $\rm{[Fe/H]}$ measurements 
and are hotter than the Grundahl jump, with \Teff $> 11,500$ K. 

\begin{figure*}
   \centering
   \includegraphics*[width=7.0in]{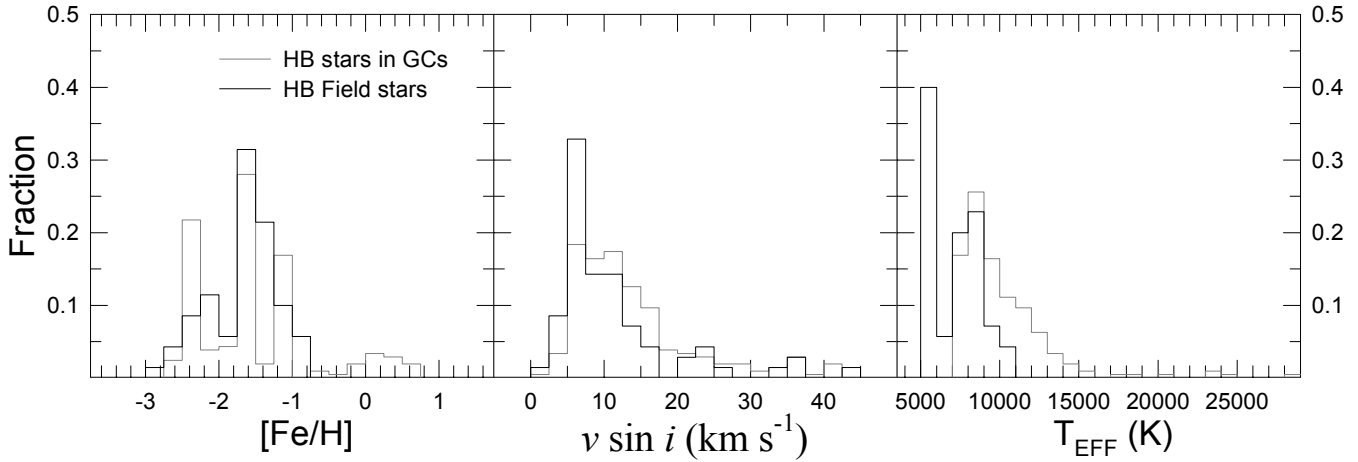}
   \caption{Histograms of [Fe/H], \vsini and \Teff for horizontal branch (HB) stars 
            (from \textit{left} to \textit{right panels}). Field HB stars 
            are indicated by \textit{black lines}, and HB stars in GCs 
            by \textit{gray lines}. While the [Fe/H] histograms 
            are very different, their \vsini distributions look very similar
			(see Fig.~\ref{HBG}).}
              \label{HB_histo}%
    \end{figure*}

Figure~\ref{HB_histo} shows the histograms of  
$\rm{[Fe/H]}$, \vsini and \Teff for stars in the field and GCs
contained in our sample. Apart from the discreet nature of the GC [Fe/H] distribution, 
there are no outstanding differences between the histograms for the field and GC samples.  
In addition, the GC distribution clearly reveals the presence of stars with 
enhanced iron abundaces, as produced by
the effects of metal levitation. As previously noted, 
HB stars with \Teff $\geq 11,500$~K are 
affected by radiative levitation and gravitational settling, 
thus the amount of iron and other heavey elements 
in their atmospheres is strongly enhanced  
(e.g., Grundahl et al. 1999; Moehler et al. 1999, 2000; Fabbian et al. 2005; 
Pace et al. 2006; and references therein). 
However, this effect cannot be noted among the field stars in our sample, because 
there are important differences in the \Teff distribution of both samples
(see Fig.~\ref{HB_histo}, \textit{right panel}). Our sample of field stars is comprised 
of blue and red HB stars, and  no stars present \Teff $> 10,500$ K, 
whereas the stars in the GC sample are blue HB stars only. To be sure, 
Behr (2003b) does list a few HB stars with higher temperatures, but they all have, 
probably as a consequence of radiative levitation, high metallicities. Since we 
are unable to infer these stars' original metallicities, we do not include them 
in our sample of field HB stars.    
In the case of GCs, the original metallicities of the HB stars showing radiative 
levitation effects can be easily determined, since they should~-- at least in 
monometallic clusters~-- be basically 
identical to those of stars in the same GCs that are not affected by radiative 
levitation. 

\begin{figure}
   \centering
    \includegraphics*[width=3.2in]{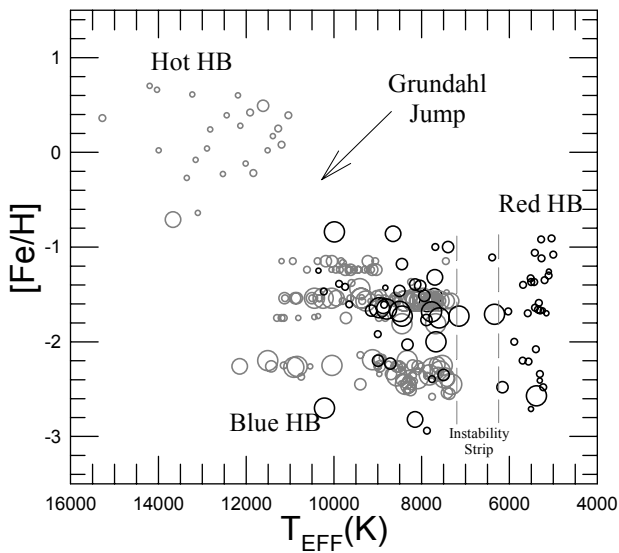}
   \caption{ \vsini values for the HB stars 
            as a function of  metallicity and \Teff. As in Figure~1, 
            the symbols represent differents intervals of \vsini.
	          Field HB stars are presented with black open symbols, whereas  
	          HB stars in  Galactic GCs are shown as gray open 
	          symbols. The differents groups of HB stars (hot HB, blue HB, red HB) 
			  and the discontinuity in 
              \vsini values and [Fe/H], produced by the Grundahl jump, are  
			  readily apparent.}
              \label{HBG}%
    \end{figure}

\begin{figure}
   \centering
    \includegraphics*[width=3.2in]{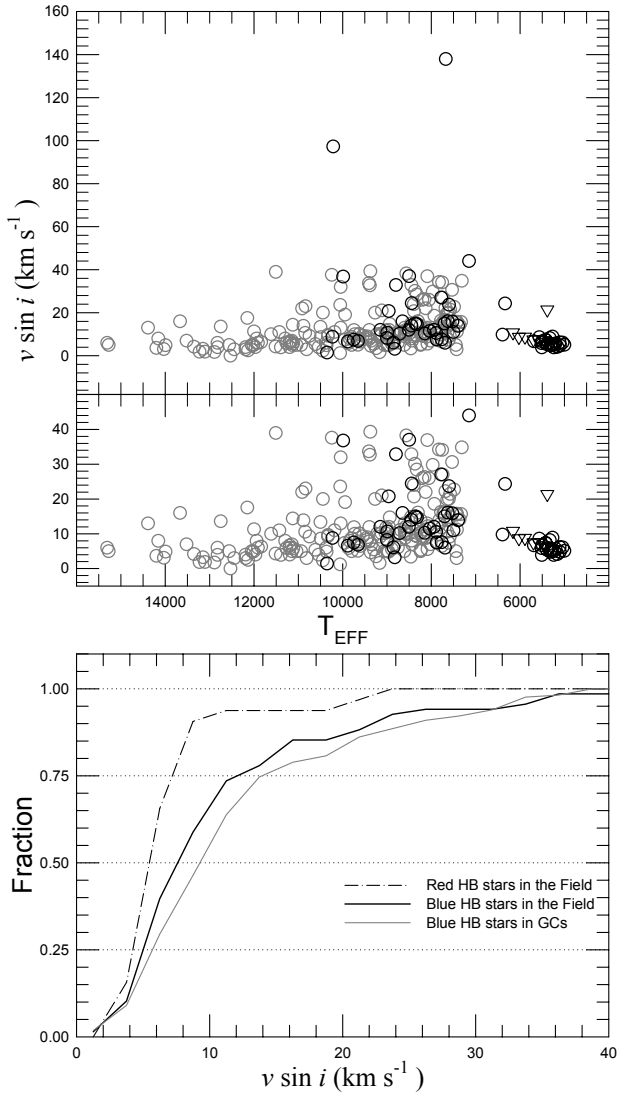}
  \caption{The \textit{upper panel}  shows  \vsini as a function of 
            \Teff for HB stars. Black circles 
            represent field HB stars,	gray circles  
            the HB stars in Galactic GCs, and the inverted triangles 
            the field red HB stars. The \textit{lower panel} 
            shows the corresponding cumulative \vsini distribution, 
			 where the two stars with unusually high rotation 
			velocities (\vsini $\gtrsim 100$~\kms) were not included; 
			see text). The solid gray line represents the distribution of blue HB 
			stars in GCs, the solid black line represents the 
            distribution of  blue HB stars in the field, and the dash-dotted 
            line represents the distribution of field red HB stars.}
              \label{HBT}%
    \end{figure}

Figure~\ref{HBG} shows the rotational velocities of the HB stars as a function of 
temperature and $\rm{[Fe/H]}$. 
We can see three different groups, 
characterized by different distributions in \vsini, $\rm{[Fe/H]}$, and 
\Teff. We can identify these   groups as the hot HB (i.e., HB stars hotter than 
the Grundahl jump), the blue HB (cooler than the jump),  and the red HB. 
The hot HB stars present \Teff $\geq 11,500$ K and high values of $\rm{[Fe/H]}$ 
(again due to radiative levitation). 
The blue HB stars present $7200~<$~\Teff~$< 11,500$~K, and the red HB stars have \Teff 
lower than $6300$~K. Also shown is the instability strip, where the RR Lyrae stars are located.
Figure~\ref{HBT} presents the values of \vsini as a function of \Teff 
(\textit{upper panel}) and the cumulative \vsini distribution (\textit{lower panel}) for the 
different groups of HB stars.
 
As proposed by  Vink \& Cassisi (2002), hot HB stars may lose mass, and such mass 
loss by  stellar winds could also lead to a loss in angular momentum and thus a reduction in 
rotation velocities, thus explaining their low rotation velocities (see Fig.~\ref{HBG}) of the 
HB stars hotter than the Grundahl jump (Sweigart 2002). 
Also, the strong element gradients in the atmospheres of these 
stars can inhibit angular momentum transport, thus also preventing these stars from 
becoming fast rotators, even if they are able to preserve a rapidly rotating core 
(Sills \& Pinsonneault 2000). 
On the other hand, the stellar wind is predicted to become 
very weak in stars with \Teff$<10000$~K.  Indeed, in Figure~\ref{HBT} we can see 
that the values of \vsini for the blue HB stars tend to be higher when \Teff$< 10000$~K, 
whereas the stars with \Teff $> 10000$~K present low \vsini values. 
Figure~\ref{HBT} also shows that the red HB stars are, mainly, 
slow rotators, thus suggesting an overall dependence of \vsini with  
\Teff (Carney et al. 2008).  

Note that there are two stars classified as HB stars
with \vsini values much higher than for 
the other fast HB rotators, namely $\rm{BD+01^{o}~0514}$ and 
$\rm{BD+30^{o}~2355}$, which present \vsini~=~137.9~\kms and 97.3~\kms, respectively. 
Behr (2003b) suggested that  
$\rm{BD+01^{o}~0514}$ is an RR Lyrae star, but to the best of our knowledge 
a variability analysis is not yet available for this star; additionally, RR Lyrae 
stars are known to be slow rotators (Peterson et al. 1996).  
$\rm{BD+30^{o}~2355}$ was, in turn, catalogued by Behr as a post-HB star.
In view of their atypical rotation velocities, we suggest that neither of these 
stars is an HB or post-HB star, being more likely MS stars. Until this issue
is resolved, these stars will not be considered in the following analysis. 

We  analyze the \vsini distribution of  
the blue (stars with $7200<$~\Teff~$<11,500~K$ in the field and the GCs)  
and red groups of HB stars (see \textit{lower panel} in 
Figure~\ref{HBT}). In these groups we found $<$\vsini$>=11.46\pm8.63,\,12.81\pm8.63,\,
7.55\pm4.38$\kms for blue HB stars in the field and the GCs, and red HB stars in the field, 
respectevely. Specifically we found  $<$\vsini$>_{\rm{FE1}}=16.03\pm14.69$\kms,
$<$\vsini$>_{\rm{FE2}}=11.91\pm9.07$\kms and $<$\vsini$>_{\rm{FE3}}=8.96\pm4.39$\kms for blue 
HB in the field, $<$\vsini$>_{\rm{FE2}}=12.22\pm7.99$\kms and $<$\vsini$>_{\rm{FE3}}=14.45\pm9.70$\kms for blue 
HB in the GCs (no blue HB stars in FE1 group), and $<$\vsini$>_{\rm{FE1}}=5.65\pm0.64$\kms,
$<$\vsini$>_{\rm{FE2}}=7.44\pm4.27$\kms and $<$\vsini$>_{\rm{FE3}}=8.17\pm5.14$\kms for red 
HB in the field. We can note that the mean values of \vsini for the blue HB star groups
are very similar. The cumulative \vsini distributions reveal differences 
between the field and  the  GCs. However, this is not a strong result, and we should 
also keep in mind the relative paucity of the field star sample, with only 36 blue HB 
stars.  We can also see that red and blue HB stars present different distributions,  
with the red HB stars presenting lower \vsini than the blue ones. 
We have applied the Kolmogorov--Smirnov (KS) test on the distributions of
\vsini of blue HB stars (field stars
and stars in GCs) . We found that  the maximum distance between both distributions
is $D_{max}=0.166$ and the probability that both distributions come from the same
 parent distribution is $P_{\rm KS}=94.5 \%$,
 thus  suggesting that  the enviroment has not played a strong role in shaping the 
 overall rotation distribution for HB stars (Behr 2003b).   
 However, this does not rule out the possibility that, for some individual GCs, the 
rotation behavior has been markedly different than for field HB stars and for other GCs 
alike.

 
 A very interesting property that is shared between blue HB stars in the field and in GCs 
 is that in both groups the fraction of fast rotators is similar: approximately $31\%$ and $25\%$ 
of the blue HB stars have 
 rotation velocities \vsini $>15.0$ \kms in the field and the GCs, respectively. 
We would like to remind the reader that the differences in the $\rm{[Fe/H]}$ distributions 
between field  and  GC stars do not affect the corresponding \vsini distribution 
 (see Fig.~\ref{HB_histo},  \textit{center panel}). On the other hand, 
the \vsini distribution of  red HB stars is again markedly different from the distributions 
for the  hot and blue HB stars, with the amount of fast rotators being only  about $6\%$  
(but again we remark that the number of stars in the red HB sample is quite small). 
Obviously, observations of enlarged samples will be needed to understand the behavior 
of rotation as a function of temperature along the HB. In this sense, it would be very 
interesting to derive rotation velocities for a large sample of RR Lyrae stars, in a 
temperature regime intermediate between the red and blue HB stars, but for which 
there is very little data currently available, suggesting however little or no rotation 
(e.g., Peterson et al. 1996).  

Recio--Blanco et al. (2002) suggested that there is no correlation between the value 
of \vsini and the evolutionary stage on the HB. If confirmed, this would have important 
implications for our understanding of angular momentum evolution in HB stars, since most 
theoretical studies suggest that angular momentum loss and transport should be important 
in explaining the lack of fast rotators among stars hotter than the Grundahl jump, and 
likewise possibly their presence among blue HB stars cooler than this limit (e.g., Sills
\& Pinsonneault 2000; Sweigart 2002). 

In order to analyze this suggestion, we have obtained, from different sources in the 
literature, the Johnson and Str\"omgren photometries for the stars
with \vsini mesuarements in the GCs NGC~288,  M~3 (NGC~5272), M~4 (NGC~6121), M~13 (NGC~6205), M~15 
(NGC~7078), M~68 (NGC~4590), M~79 (NGC~1904), and  M~92 (NGC~6341) (see table~\ref{tab:GCStars}).
We have derived the absolute magnitude in both photometric systems using the 
distance modules and reddening values compiled in Harris (1996). For the case of M~79, the 
absolute magnitude    \textit{u}  and the unreddened colors $(u-y)_{0}$ in the 
Str\"omgren system were calculated using a $(m-M)_{u}=15.6$ and $E(u-y)=1.89 E(\bv)$ 
(Crawford \& Mandwewala 1976; Clem et al. 2004). We have transformed the evolutionary 
tracks and ZAHB sequences to the observational planes following the same procedures 
already described in Catelan et al. (2004) and Cort\'es \& Catelan (2008). The 
evolutionary models used in the present study are the same as those computed by 
Catelan et al. (1998) and Sweigart \& Catelan (1998). 
We have chosen the evolutionary tracks for different 
masses with metal abundances $Z = 0.0020$, $0.0010$, and $0.0005$ 
(with MS helium abundance $Y_{\rm MS}=23\%$), 
since these values of $Z$ adequately cover the metallicites of the GCs considered in our analisys.
To transform these $Z$ values into $\rm{[Fe/H]}$, we have used 
equation~(2) in Cort\'es \& Catelan (2008), based on the scaling relation of Salaris et al. (1993)
for an $\rm{[\alpha/Fe]=0.3}$, as typically found among halo stars (e.g., Pritzl et al. 2005 and
references therein).
Specifically,  the evolutionary tracks with $Z = 0.0020$ \footnote{The masses 
of the evolutionary tracks for $Z = 0.0020$ are $0.500$, $0.510$, $0.520$, $0.530$, $0.540$, $0.550$, 
$0.560$, $0.580$, $0.600$, and $0.620$ $\rm{M_{\odot}}$} were used for the GCs NGC~288 and M4,
 the evolutionary tracks with $Z = 0.0010$ \footnote{The masses of the 
evolutionary tracks for $Z = 0.0010$ are $0.495$, $0.497$, $0.506$, $0.515$, $0.527$, $0.542$, $0.558$, 
$0.575$, $0.589$, $0.604$, $0.616$, and $0.630$ $\rm{M_{\odot}}$}  for the GCs M~3, 
M~13 and M~79, and the evolutionary tracks with $Z = 0.0005$ \footnote{The masses of 
the evolutionary tracks for $Z = 0.0005$ are $0.498$, $0.499$, $0.501$, $0.508$, $0.519$, $0.531$, $0.565$, $0.600$, $0.630$, $0.660$, and $0.693$ $\rm{M_{\odot}}$} for the GCs M~92, M~15 
and M~68. No attempt was made to properly model the stars hotter than the Grundahl jump at 
11,500~K; the reader is thus warned that these models cannot be used to reliably describe
the $u$-band magnitudes in particular of these stars (Grundahl et al. 1999). 
 
\begin{figure*}
   \centering
    \includegraphics*[width=7in]{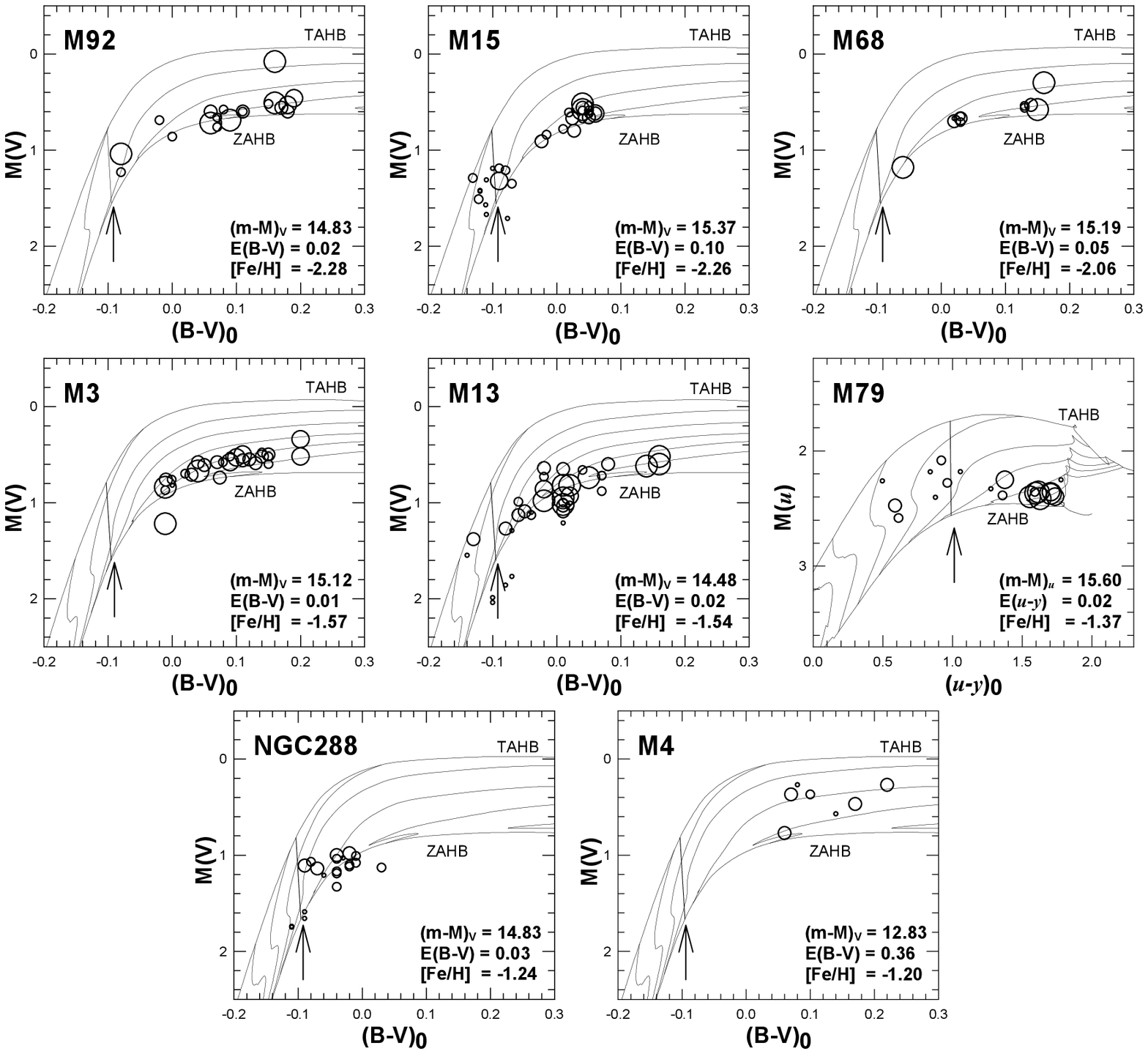}
   \caption{Color--magnitude diagram  in the Johnson and the 
            Str\"omgren  systems for HB stars 
            in the Galactic GCs M~92, M~15, 
            M~68, M~3, M~13, M~79, NGC~288, and M~4. 
            HB stars with \vsini measurements are shown  
            as open circles, the sizes of the circles 
   					representing different intervals of \vsini (see Fig.~1). 
					The ZAHB, the TAHB, and evolutionary tracks for several different mass 
					values are also shown (see text). The locus of the Grundahl jump 
					is presented in each panel using a black arrow, and in each panel the points 
					in the evolutionary tracks with $T_{\rm eff} = 11,\!500$~K 
					(Grundahl et al. 1999) are connected with a line.
				}  
              \label{HBtracks}%
    \end{figure*}

Figure~\ref{HBtracks} shows  the stars with measurements of \vsini in the different
GCs, with the ZAHBs, TAHBs (terminal-age HBs, or helium core exhaustion locus), 
and the referred evolutionary tracks overplotted. We can see that
there is no strong evidence that \vsini is linked with the evolutionary stage of the stars.
Note, in particular, in several clusters, perhaps most notably M13 and M92, we clearly find 
stars with high \vsini values very close to the ZAHB. Conversely, in several clusters 
we can also see slow rotators close to the ZAHB. This suggests that either these stars 
arrived at the ZAHB displaying rotation rates very similar to their current values, or else
that angular momentum loss and redistribution may operate extremely fast once the stars reach 
the ZAHB. 

In summary, we find that the values of \vsini for the HB stars are not correlated with 
the evolutionary distance from the ZAHB. However, enlarged samples of low-metallicity 
HB stars with accurately measured \vsini values are needed to put this result on a 
firmer footing.

%

%

%


%


\section{Conclusions}\label{sec:Conclusions}
In this paper, we have performed a careful compilation of \vsini values for metal-poor
stars ($[{\rm Fe/H}] < -0.5$) covering different evolutionary stages (MS, RGB, HB) from 
the literature. Our sample includes stars both in the field and in Galactic GCs. We also
gathered metallicity and photometric data for these stars, and present all data in the 
form of extensive tabulations. We have conducted a preliminary analysis of these data, 
and our main conclusions are as follows. 

The distribution of \vsini in the H-R diagram shows that the slow rotators are 
distributed in all evolutionary stages, from the MS to the HB. 
The fast rotators are concentrated in the HB,  suggesting 
that the HB stars somehow acquired angular momentum in the previous phase (the RGB), 
or else that RGB stars preserved rapidly spinning cores and were later, either during
the pre-ZAHB phase or on the HB proper, able to transfer angular momentum to the 
stellar surface. 
However, an analysis of evolutionary differences in \vsini values 
along the HB phase for eight different GCs reveals little or no dependence of \vsini on 
evolutionary phase, thus suggesting that 
any angular momentum transport or losses in the HB phase must either operate very 
quickly close to the ZAHB, or be very inefficient during the HB phase. 

HB stars in the field and in GCs do not reveal important differences 
in their rotational behavior, thus suggesting that the environment does not affect
the rotation behavior of these stars in an important way. However, we cannot rule out
the possibility that some individual GCs will have a peculiar stellar rotation behavior,  
driven by environmental effects. In addition, sample sizes remain relatively small, 
and it would certainly be important to derive \vsini for enlarged samples of stars, 
in the field and in GCs alike, to put these results on a firmer footing. It would 
particularly interesting to derive \vsini for a large sample of RR Lyrae stars, for
which very few measurements are currently available, and yet there is a puzzling 
indication of very little (or no) rotation (e.g., Peterson et al. 1996). 

We also find that, while our field and GC samples have markedly different metallicity 
distributions, such differences are not reflected upon marked differences in their 
corresponding \vsini distributions. It thus appears that metallicity is not a 
relevant parameter affecting the overall \vsini distribution, at least in the 
low-metallicity regime studied in this work. 

RGB stars in the Carney et al. (2008) sample reveal some intriguing evidence of 
spinning up as they approach the RGB tip. However, when we incorporate data for 
bright RGB stars from other sources, we do not find any significant variation 
in \vsini with evolutionary stage along the RGB. 

%

\acknowledgments
This work has been supported by continuous grants from the CAPES, CNPq and 
FAPERN Brazilian agencies. We are particularly grateful to CONICYT and 
CNPq for CONICYT-CNPq PCCI grant 018/DRI/061. M.C. thanks the John Simon 
Guggenheim Memorial Foundation for a Guggenheim Fellowship.

\vfill\eject

\LongTables
\begin{center}
\begin{longtable*}{lrccrcccc}
\caption[Feasible triples for a highly variable Grid]{\\Physical parameters 
and Photometry of Field Metal-Poor Stars} \label{tab:FieldStars} \\

\hline \hline \\[-2ex]
\multicolumn{1}{l}{Star ID} &
\multicolumn{1}{c}{\Teff} &
\multicolumn{1}{c}{$\log (g)$} &
\multicolumn{1}{c}{\vsini} &
\multicolumn{1}{c}{$\rm{[Fe/H]}$} &
\multicolumn{1}{c}{$V$} &
\multicolumn{1}{c}{$(\bv)$} &
\multicolumn{1}{c}{$(V-I)$} &
\multicolumn{1}{c}{Reference} \\[0.5ex] 
        &	(K)	&	(cgs)	&	(\kms)	& (dex) &	(mag)	&	(mag)	&	(mag)	&	    	\\

	\hline
   \\[-1.8ex]
\endfirsthead

\multicolumn{3}{c}{{\tablename} \thetable{} -- Continued} \\[0.5ex]
  \hline \hline \\[-2ex]
\multicolumn{1}{l}{Star ID} &
\multicolumn{1}{c}{\Teff} &
\multicolumn{1}{c}{$\log (g)$} &
\multicolumn{1}{c}{\vsini} &
\multicolumn{1}{c}{$\rm{[Fe/H]}$} &
\multicolumn{1}{c}{$V$} &
\multicolumn{1}{c}{$(\bv)$} &
\multicolumn{1}{c}{$(V-I)$} &
\multicolumn{1}{c}{Reference} \\[0.5ex] 
        &	(K)	&	(cgs)	&	(\kms)	& (dex) &	(mag)	&	(mag)	&	(mag)	&	    	\\

	\hline
  \\[-1.8ex]
\endhead

  \multicolumn{3}{l}{{Continued on Next Page\ldots}} \\
\endfoot

\hline \hline 
 \\[-0.5ex]
\multicolumn{1}{l}{REFERENCES:}\\
\multicolumn{1}{l}{(1) Kinman et al. 2000.}\\
\multicolumn{1}{l}{(2) Behr 2003a.}\\
\multicolumn{1}{l}{(3) Carney et al. 2003.}\\
\multicolumn{1}{l}{(4) De Medeiros et al. 2006.}\\
\multicolumn{1}{l}{(5) Carney et al. 2008.}\\
  \\[-1.8ex] 
\endlastfoot

\object{HD20}	&	5350	&	2.50	&	5.9	& $	-1.66	$&	9.07	&	0.54	&	0.75	&	3	\\
\object{HD97}	&	5270	&	2.83	&	4.0	& $	-1.42	$&	9.40	&	0.64	&	0.82	&	2	\\
...	&	4910	&	2.60	&	4.0	& $	-1.38	$&	9.40	&	0.64	&	0.82	&	3	\\
...	&	4953	&	3.10	&	4.2	& $	-1.40	$&	9.40	&	0.64	&	0.82	&	4	\\
\object{HD2857}	&	7550	&	3.00	&	29.0	& $	-1.70	$&	9.95	&	0.17	&	0.24	&	1	\\
...	&	8002 &	3.38	&	25.1	& $	-1.67	$&	9.95	&	0.17	&	0.24	&	2	\\
\object{HD2880} &	4810	&	4.01	&	6.5	& $	-0.83	$&	8.52	&	0.84	&	0.88	&	2	\\
\object{HD3008}	&	4140	&	1.00	&	9.2	& $	-1.43	$&	9.70	&	1.14	&	1.23	&	3	\\
\end{longtable*}
\end{center}
%

\vfill\eject
\begin{center}
\begin{longtable*}{lrcrcccrccc}

\caption[Feasible triples for a highly variable Grid]{\\Physical parameters 
and Photometry of Metal-Poor Stars in Globular Cluster}\label{tab:GCStars} \\

\hline \hline \\[-2ex]
\multicolumn{1}{l}{Star ID} &
\multicolumn{1}{c}{\Teff} &
\multicolumn{1}{c}{$\log (g)$} &
\multicolumn{1}{c}{\vsini} &
\multicolumn{1}{c}{$\rm{[Fe/H]}$} &
\multicolumn{1}{c}{$V$} &
\multicolumn{1}{c}{$(\bv)$} &
\multicolumn{1}{c}{OTHER} &
\multicolumn{1}{c}{$M_{V}$} &
\multicolumn{1}{c}{$(\bv)_0$} &
\multicolumn{1}{c}{Reference} \\[0.5ex] 
        &	(K)	&	(cgs)	&	(\kms)	& (dex) &	(mag)	&	(mag)	&	(mag)	&	 (mag) & (mag) &   	\\

	\hline
   \\[-1.8ex]
\endfirsthead

\multicolumn{3}{c}{{\tablename} \thetable{} -- Continued} \\[0.5ex]
  \hline \hline \\[-2ex]
\multicolumn{1}{c}{Star ID} &
\multicolumn{1}{c}{\Teff} &
\multicolumn{1}{c}{$\log (g)$} &
\multicolumn{1}{c}{\vsini} &
\multicolumn{1}{c}{$\rm{[Fe/H]}$} &
\multicolumn{1}{c}{$V$} &
\multicolumn{1}{c}{$(\bv)$} &
\multicolumn{1}{c}{OTHER} &
\multicolumn{1}{c}{$M_{V}$} &
\multicolumn{1}{c}{$(\bv)_0$} &
\multicolumn{1}{c}{Reference} \\[0.5ex] 
        &	(K)	&	(cgs)	&	(\kms)	& (dex) &	(mag)	&	(mag)	&	(mag)	&	 (mag) & (mag) &   	\\

	\hline
  \\[-1.8ex]
\endhead

  \multicolumn{3}{l}{{Continued on Next Page\ldots}} \\
\endfoot

\hline \hline 
 \\[-0.5ex]
 

\multicolumn{1}{l}{REFERENCES:}\\
\multicolumn{1}{l}{(1) Peterson 1985.}\\
\multicolumn{1}{l}{(2) Peterson et al. 1995.}\\
\multicolumn{1}{l}{(3) Cohen 1997.}\\
\multicolumn{1}{l}{(4) Recio--Blanco et al. 2002.}\\
\multicolumn{1}{l}{(5) Behr 2003a.}\\
\multicolumn{1}{l}{(6) Crocker et al. 1988.}\\
\multicolumn{1}{l}{(7) Moehler et al. 1995.}\\
\multicolumn{1}{l}{(8) Fabbian et al. 2005.}\\

\\[-1.8ex] 
\endlastfoot
%
M3/19	&	7607	&	3.01	&	7.40	&	...	&	15.64	&	0.16	& $	...	$ &	0.52	&	0.15	&	2	\\
M3/20	&	7640	&	3.02	&	10.60	&	...	&	15.64	&	0.15	& $	...	$ &	0.52	&	0.14	&	2	\\
M3/54	&	7764	&	3.05	&	11.80	&	...	&	15.67	&	0.13	& $	...	$ &	0.55	&	0.12	&	2	\\
M3/97	&	7579	&	3.01	&	14.80	&	...	&	15.62	&	0.16	& $	...	$ &	0.50	&	0.15	&	2	\\
M3/109	&	7583	&	3.01	&	9.20	&	...	&	15.60	&	0.15	& $	...	$ &	0.48	&	0.14	&	2	\\
M3/168,1371,AX	&	7702	&	3.04	&	16.00	&	...	&	15.62	&	0.12	& $	...	$ &	0.50	&	0.11	&	2	\\
M3/201,1130,II-63	&	7818	&	3.07	&	12.70	&	...	&	15.68	&	0.12	& $	...	$ &	0.56	&	0.11	&	2	\\
M3/237	&	7932	&	3.09	&	18.60	&	...	&	15.70	&	0.10	& $	...	$ &	0.58	&	0.09	&	2	\\
\end{longtable*}
\end{center}
%

%

\end{document}